\documentclass[a4paper, margin=2cm, 10pt]{article}

\usepackage[backgroundcolor=blue!10!white]{todonotes}
\usepackage[top=2in, bottom=1.5in, left=1.5in, right=1.5in]{geometry}

\usepackage[adobe-utopia]{mathdesign}
\usepackage{amsmath}
\usepackage{enumitem}
\usepackage{graphics}
\usepackage{color, colortbl, soul}

\definecolor{darkBlue}{RGB}{0,0,205} 

\usepackage[colorlinks]{hyperref}
\hypersetup{
    allcolors={darkBlue},
}

\usepackage[natbib=true,style=authoryear-icomp,sortcites=false,firstinits=true,maxcitenames=2, maxbibnames=99,useprefix=true,dashed=false,url=false,doi=false,isbn=false,eprint=false,backend=bibtex8]{biblatex}

\bibliography{bib,him}

\usepackage{longtable}
\usepackage{afterpage}
\usepackage{tabularx}
\usepackage[singlelinecheck=off]{caption}
\usepackage{array}
\usepackage{rotating}
\usepackage{booktabs}

\providecommand{\keywords}[1]{\textbf{\textit{Keywords:}} #1}

\title{
Information Verification for Humanitarians: \\
A Critical Review
}

\author{Yilin Huang\\
	Delft University of Technology\\
	\href{mailto:y.huang@tudelft.nl}{y.huang@tudelft.nl} 
	\and 
	Christophe Billen \\
	People's Intelligrence\\
	\href{mailto:c.billen@peoples-intelligence.org}{c.billen@peoples-intelligence.org} 
	}

\date{June 2018}

\begin{document}

\maketitle

\begin{abstract}
Quality humanitarian information is essential for efficient, effective and coordinated humanitarian responses. During crises, however, humanitarian responders rarely have access to quality information in order to provide the much needed relief in a timely fashion. Traditional methods for the acquisition and evaluation of humanitarian information typically confront challenges such as  poor accessibility, limited sources, and the capacity of monitoring and documentation. The more recent emergence of user generated content from online social platforms addressed some challenges faced by traditional methods, but it also raised many concerns regarding information quality and verifiability, among others, that affect both the public and humanitarian actors. This paper provides an overview of information verification methods in literature  and reviews information collection and verification practices and tools used by news agencies and humanitarian organizations. Twenty crowd-sourced information projects in humanitarian and human rights nature are surveyed. We discuss the findings and give recommendations for future research.

\end{abstract}

\small{\keywords{information verification, user generated content, source evaluation, cross-validation}}

\tableofcontents

\section{Introduction}
\label{sec:intro}

Since the 1950s, the number and magnitude of disasters have increased exponentially \citep{Ozdamar2015}. About 300 million people on the average are effected annually since the 1990s \citep{Ozdamar2015}. 
Faced with complex humanitarian situations, responders rarely have access to quality information for decision-making to provide much the needed relief in a timely fashion. 
Besides the local knowledge, examples of such information include the aid requests of victims, the numbers and locations of internally displaced persons, the incidents reports, the conditions of essential local infrastructures, just to name a few. Humanitarian information is valued as the \textit{sine qua non} of humanitarian response \citep{OCHA2006}, and humanitarian information management and exchange are the principle source of situational awareness, crisis decision-making and coordination \citep{Altay2014}. 

Traditional methods for the acquisition and evaluation of humanitarian information typically confront a number of challenges. 
First, to deliver aid and assistance, humanitarian actors need to assess the situation and/or to seek representatives of the affected populations for interviews, but the actors are not always present in the vicinity of conflict or disaster zones, or have access to those areas. 
Second, due to limited resources, accessibility, security and time, it is often very hard if not impossible to find victims and witnesses who can and are willing to provide potentially sensitive information. 
Third, humanitarian actors rarely have the capacity to continually monitor and document the incidents over time in the affected areas, and to provide the affected populations timely and effective aid in return.



In recent humanitarian crises such as the 2010 Haiti earthquake and the 2011 Egyptian revolution, incorporating humanitarian information from social media and User Generated Content (UGC) proved useful when the information was inspected at an aggregated level 
\citep{Dugdale2012}. 
During those crises, online platforms such as Twitter and Facebook facilitated reporting information more efficiently than traditional communication channels \citep{Norheim2010,Hermida2014}, and addressed some challenges faced by the traditional methods.
The information contained in UGC can be vital for effective response \citep{Takahashi2015}, and can be used to boost the speed and accuracy of relief operations in real-time, and to empower and uplift the morale of the local populations \citep{Carley2016,Conrado2016,Panagiotopoulos2016,Haworth2016}. 
Nevertheless, existing humanitarian information systems using technologies such as social media and crowd-sourcing have shortcomings including but are not limited to the following \citep{Tapia2011,Dugdale2012,Haworth2016, Conrado2016, Anson2017}:  

\begin{enumerate}
\item The systems are not effective in collecting relevant and quality information. While there is information overload and processing difficulties, there is also a high risk of receiving inaccurate and incorrect information (including that from malicious users). 
\item There is no or limited evaluation of the reliability of the sources and the credibility of the information, making humanitarian actors and affected communities vulnerable to inaccurate and incorrect information. 
\item The  contributed information has largely been deemed as unverifiable and untrustworthy. Thus it is construed  as  unsuitable  to  incorporate  into  established 
mechanisms  for  decision-making.
\ 
\item There is a lack of feedback loops and empowerment of those (often the affected populations) who contributed the information, partly due to the above shortcomings.
\end{enumerate}

The quality of information from UGC is a major challenge that affects both the public and humanitarian actors \citep{Haworth2016}. The vast volume of UGC circulating in social media contains relevant and useful information, which is potentially life-saving, but it also contains floods of irrelevancy, inaccuracy and rumours  \citep{Anson2017}. For these reasons, although there have been needs and interests from the humanitarian actors and local communities in the field to establish (effective and reliable) information exchange, many humanitarian actors are sceptical about the levels of reliability of self-reported information since the information is often unverifiable \citep{Altay2014,Conrado2016}. 

How to detect relevant humanitarian information and verify the information in an effective and efficient manner is the concern of this paper. 
In the following, we first provide an overview of information verification methods in literature, then review 
information collection and verification practices and tools used by news organizations and humanitarian organizations. 
We end with a discussion of the findings and recommendations for future research.

\section{An Overview of Information Verification Methods}
\label{sec:methods}

In literature, some researchers make no distinction between data (quality) and information (quality) \citep[e.g.,][]{Wand1996, Pipino2002, Loshin2011} while some others see the difference as being crucial \citep[e.g.,][]{Ackoff1989, Lillrank2003,Price2005}. The definitions of data or information quality in literature are also equivocal. In this paper,  information quality refers to the semantic and pragmatic clarity of UGC rather than its syntactic clarity \citep{Huang2013}. 
Important dimensions of information quality include relevance, accuracy, volume, completeness, timeliness, reliability and verifiability \citep{SHANTEAU199275,yildiz2015handbook,shamala2017integrating}. In relation to those, veracity can be defined as the combination of how accurate, complete, reliable and timely the information in question is \citep{lin2016}. Veracity can also include trustworthiness which is an aggregated dimension determined by the data origins (or the sources), and the data collection and processing methods \citep{lin2016}.

Information detection and verification are often researched in the context of investigative journalism and (business, police, civilian or military) intelligence, and in more general applications. 
Four major types of (not necessarily mutually exclusive) (text-based) information detection and verification methods  can be identified in literature: 1) cross-validation, 2)~expert opinion, 3)~crowd-sourcing, and 4) machine learning; see Table \ref{tab:methods}.

\begin{description}
	\item[Cross-validation] a.k.a. triangulation, of independent data sources is the process where humanitarian actors utilize additional information to validate the veracity of given information extracted from UGC \citep{crowley2013decision}. A major limitation is the required manual input of users for validation. Its effectiveness is directly and entirely dependent on the skill and ability of the users \citep{DAUME20149}.
	\item[Expert opinion]  is the process where experts or people of authority utilize their expertise or authoritative sources to validate the veracity of information \citep{Martin2016}. This type of methods is limited by the availability of experts in the field \citep{Martin2016}. Similar to cross-validation, it often requires extensive manual input, and the knowledge, skill and network of the users \citep{Martin2016}. 
	\item[Crowd-sourcing]  is the use of Internet platforms in combination with the input of social media in order to validate the veracity of information harnessed from UGC \citep{Riccardi2016}. The users verify whether the given information is of good quality. This 
	requires a large number of users to determine the veracity of the information (a.k.a the wisdom of the masses, or collective intelligence) \citep{Howe2008,Basu2016}. 
	\item[Machine learning]  is the technology of getting computer systems to act without being explicitly programmed (\cite{michalski2013machine}), achieved through automated statistical methods (\cite{alpaydin2014introduction}). Machine learning is applied in many fields such as voice and image recognition, financial predictions, information verification and many other fields. Decision makers in the humanitarian domain are still hesitant to use such methods due to the uncertain accuracy and poor understandability \citep{Altay2014,Conrado2016}. 
	
\end{description}

\begin{table}[h!]\small
\resizebox{\columnwidth}{!}{%
\def\arraystretch{1.2}
\begin{tabular}{m{2cm}>{\raggedright\arraybackslash}m{6cm}m{4.5cm}} \arrayrulecolor{gray!40}\hline
\cellcolor{gray!25} \textbf{Method} & \cellcolor{gray!25} \textbf{Limitation} & \cellcolor{gray!25} \textbf{Reference}\\ \hline 
Cross-validation  &
\begin{itemize}[noitemsep,leftmargin=*,topsep=15pt]
  \item Restricted by the required user input 
  \item Directly and entirely dependent on users' skill and ability
  \item No or poor detection of malicious users and rumours 
  \end{itemize}
&
\begin{itemize}[label={},noitemsep,leftmargin=*,topsep=15pt]
  \item \citet{crowley2013decision}
  \item \citet{DAUME20149}
\end{itemize} \\ \hline
Expert Opinion 
&
\begin{itemize}[noitemsep,leftmargin=*,topsep=15pt]
	\item Limited availability of experts
	\item Restricted by the required user input 
	\item Dependent on users' knowledge, skill and network
	\item No or poor detection of malicious users and rumours 
  \end{itemize}
&
\begin{itemize}[label={},noitemsep,leftmargin=*,topsep=15pt]
  \item \citet{Martin2016}
\end{itemize}\\ \hline
Crowd-sourcing &
\begin{itemize}[noitemsep,leftmargin=*,topsep=15pt]
	\item Requires a great number of users 
	\item Restricted by the required user input 
	\item No or poor detection of malicious users and rumours 
  \end{itemize}
&
\begin{itemize}[label={},noitemsep,leftmargin=*,topsep=15pt]
  \item \citet{Basu2016}
  \item \citet{Callaghan2016}
  \item \citet{Gao2011}
   \item \citet{Ludwig2017}
  \item \citet{Meier2011}
  \item \citet{Riccardi2016}
  \item \citet{Soden2014}
   \item \citet{Yuan2018}
\end{itemize}\\ \hline
 Machine Learning
 &
\begin{itemize}[noitemsep,leftmargin=*,topsep=15pt]
  \item Needs good training data
  \item Uncertain accuracy rates 
  \item Hard to obtain trust from users because of its untransparent inner working
  \item No or poor detection of malicious users and rumours 
  \end{itemize}
&
\begin{itemize}[label={},noitemsep,leftmargin=*,topsep=15pt]
   \item \citet{Ali2017}
   \item \citet{Carley2016}
   \item \citet{Castillo2013}
   \item \citet{Diakopoulos2012}
   \item \citet{Hung2016}
   \item \citet{Kang2012}
   \item \citet{Liu2016}
   \item \citet{Ozdamar2015}
   \item \citet{Spence2016}
\end{itemize}\\ \hline
\end{tabular}
}
\caption{Four major types of information detection and verification methods \citep{Vaporidis2019}}
\label{tab:methods}
\end{table}

In addition, all the above methods do not explicitly and effectively detect malicious uses and rumours along the detection and verification of relevant information. 
Malicious uses and rumours are sources of instability during relief operations \citep{Conrado2016,Riccardi2016}. They can disrupt the flow of humanitarian operations \citep{Haworth2016}. For example, terrorists and kidnappers, and sometimes even the affected communities, media and humanitarian actors might publish false or unverified information  \citep{Altay2014,Riccardi2016}. The abuse and misuse of information can create additional conflict and problems, and possibly put people in danger and jeopardize the success of the relief operation  \citep{Riccardi2016}.

\section{Information Verification in News Organizations}

Some news agencies (particularly investigative journalists) and humanitarian actors lately started using information verification tools for UGC  due to the emergence of new media \citep{Brandtzaeg2016,Altay2014}: 

\begin{itemize}
\item News organizations traditionally focuses on ``breaking news''. Some now focus more on being the best at verifying
and curating the information \citep{Newman2009}. 
\item Established humanitarian actors traditionally operate  with  centralized  command  structures,  standard  operating  procedures, and internal   vetting   standards to ensure the flow, accuracy and verification of information. With the current expectation of speed and efficiency, there are transitions toward harvesting UGC in combination with verification \citep{Coyle2009, Tapia2011, Walton2011}. 
\end{itemize}

In the following, we review the verification practices and tools reported in literature. 
``Practices'' refers to the information verification processes, methods and techniques through which the content is verified. ``Tools'' are for instance the
computerized or manual checklists, and software platforms that facilitate the
verification.

Established news agencies have their own verification practices, or outsource the verification of sources to other companies \citep{Hermida2014,Schifferes2014}. There are three common practices \citep{Bruno2011}:

\begin{enumerate}
\item The centralized approach, that tries to verify information within one's own organization. 
\item The decentralized approach, that tries to incorporate the crowd in verification through live blogging or streaming. 
\item The community-based approach, which tries to create a platform for
verification through crowd-sourcing. 
\end{enumerate}

For example, the \textit{BBC} created its centralized \textit{UGC Hub} in 2005, \textit{The Guardian} has its decentralized verification platform, and the \textit{CNN} has its community-based \textit{iReport} launched in 2006 \citep{Bruno2011}. 
At the UGC Hub of BBC, potentially valuable photos, texts and emails are verified before they
are published. This is done centrally at the BBC news room, by calling the
contributor personally, when possible, and asking basic questions regarding the authenticity of the content \citep{Harrison2010}. 
The content is subsequently verified by cross-validation \citep{Popoola2013}. 
For photographs, the precise place and time of the photos are
important factors for verification. They are compared and verified with the statement of the source \citep{Bruno2011}. 
The UGC Hub uses four metrics to verify the credibility of Twitter accounts \citep{Popoola2013}:
(1) the number of Twitter followers an account has; 
(2) is the account followed by an reputable source?
(3) previous posts by the account;  and
(4) how long the account has been active for?
For emails the IP addresses are checked, and for phone calls, the number prefix. Nonetheless, reaching out to
the sources in person remains the most important verification method at the BBC UGC Hub \citep{Bruno2011}. 
At CNN, a different approach is used focusing on the online community of
contributors. The information posted on \textit{iReport} is not checked prior to publication, but can be
verified by other users afterwards. Readers can recognize verified stories by badges given
when the stories are verified by other users \citep{Bruno2011}. 

Another example of community-based approach is the \textit{U-Shahid} project. It is launched by an Egyptian group based in Cairo worked with a journalist from \textit{Thomson Reuters}. The project developed a checklist during the 2011 Egyptian crisis,
with four principles for the verification of sources \citep{Meier2015}:
(1) unknown sources are called back when possible. The source is asked if he or she was an eyewitness, or if more information can be provided; 
(2) a trusted source in the area is contacted for verification when possible, and
trusted NGO workers are contacted; 
(3) online research is performed, to look for similar videos, photographs and blog posts; and
(4) cross-validation of information with reports received.

During the Arab spring, Andy Carvin, a journalist of \textit{National Public Radio} (NPR), asked his Twitter followers to verify reports \citep{Hermida2014,Meier2015}. He received news, and verified
it by retweeting it and asked for eyewitnesses and sources. They helped him translate,
cross-validate and track down the key information \citep{Silverman2014}. Unreliable sources were dropped 
and reliable accounts were saved \citep{Hermida2014}. 

News agencies such as \textit{Al Jazeera}, \textit{The New York Times} and \textit{The Wall Street Journal} have outsourced the verification of photo and video content to \textit{Storyful}, a company founded in 2010 and bought over by \textit{News Corp} for \$25 million in 2013 \citep{Hermida2015}. Storyful verifies Twitter sources by looking at:
(1)~the time of day of posts; 
(2) the weather in the content versus the actual weather reports; 
(3) the accents spoken in video content; and 
(4) the landmarks which can be confirmed by other sources \citep{Popoola2013}. 

There are a few other tools developed for information verification used by journalists. For example, \textit{Tweetdeck} is a tool for checking, and screening Twitter posts, also used by media
organizations \citep{Sump2012}. The \textit{Reuters} created its own computerized tool for verifying real-time news events on Twitter. All its event processing is computerized, and
machine learning is used to verify fake news Tweets \citep{Liu2016}.
Besides those, tools such as \textit{TinEye} and \textit{Google reverse image search} are also used for the verification of photos \citep{Pantti2015}. The EU \textit{Social Sensor project}  aims at creating groups of reliable Twitter users to verify
posts and accounts, and designing a new tool to search across social media 
for news stories, surface trends, and help with verification \citep{Schifferes2014}.


\section{Information Verification in Humanitarian Organizations}

The basis of information verification by humanitarian organizations lies in
the principles of Humanitarian Information Management and Exchange endorsed by OCHA (United Nations Office for
the Coordination of Humanitarian Affairs) and many other humanitarian actors \citep{VandeWalle2008}. One of the principles -- verifiability -- stresses the ability to ensure that information represents what it supposed to represent and the methodologies to validate information are sound; another principle -- reliability -- stresses
the credibility of the source and the method of collection.

The verification of information from UGC in humanitarian situations can take an intrinsic and/or extrinsic approach \citep{Conrado2016}. With an intrinsic approach, the validity of the content, context and the contributor themselves (i.e. the intrinsic properties) are researched. With an extrinsic
approach, additional resources to validate the information (i.e. the extrinsic properties) are searched, e.g., experts, crowd-souring and linked data \citep{Conrado2016}. 

The idea behind crowd-sourcing is that although ``truth'' is uncertain, with enough volume, a ``truth'' emerges that diminishes false
reports \citep{Okolloh2009}.
Four elements can be checked to confirm crowd-sourced information in humanitarian context: 
1)~Provenance: confirm the authenticity of the piece of information. Is this the original piece of content?
2) Source: confirm the source. Who uploaded the content?
3) Date: confirm the date of the event and the time of the content. When was the content created?
4) Location: confirm the geolocation. Where was the content created? \citep{Wardle2014}.

In the aftermath of the 2010 Haiti earthquake, UGC contributed to humanitarian responses. It is led by the \textit{Ushahidi} platform and street mapping platforms such as the
\textit{Humanitarian OpenStreetMap Team} (HOT) \citep{Soden2014}. 
%
These platforms were used because 
 traditional information gathering in humanitarian organizations were not designed to integrate intelligence from local communities, and  individual communications of the Haitians
were lost \citep{Heinzelman2010}.
A challenge during the Haiti earthquake response was to make a reliable crisis map of the affected
area for humanitarian actors to focus their relief efforts on  \citep{Meier2011}.
\textit{Ushahidi-Haiti} provided crisis mapping created by a team of volunteers based on reports received
via Twitter, email, SMS and other sources \citep{Norheim2010}. Information verification at the
Ushahidi platform can be performed in two ways. On
the site of Ushahidi, there is a verification button that allows the crowd to verify the content of the crises map \citep{Gao2011}.
But generally only a few people served as verifiers across Ushahidi map cases \citep{Gao2011}. 
The second way is to manually check the reports by Ushahidi staffs. The approval  process was rather ad hoc\footnote{``Where possible, we
called or emailed reporters to try to verify reports. Where
people reported anonymously, stories were counter-checked
by comparing with other sources e.g. mainstream media.
Where information appeared credible but we could not verify
it, we posted it and noted that it was not verified'' \citep{Okolloh2009}.} which is a risk with any crowd-sourcing tool \citep{Okolloh2009}.

\textit{Verily} is a platform designed for rapid collection and assessment of information generated during natural disasters \citep{Popoola2013}. 
The  departure  point is  the  posting  of  a  verification  request, structured as a yes/no event-based question, e.g., ``Is the Brooklyn bridge damaged in the storm?''
The request shall trigger the collection of evidence to assess, through evaluation of collected evidence, whether a given event has actually happened.  
There are also initiatives that aim to create networks of volunteers worldwide to help verify humanitarian information during crises, and to create Crisis Maps, e.g.  CrisisCommons, CrisisMappers, Standby Task Force, Humanity
Road \citep{Ziemke2012,Rogstadius2013,Cobb2014,Norris2017}. 
%

Table \ref{tab:projects}  provides a list of projects that use crowd-sourcing to acquire information in humanitarian and human rights nature\footnote{We used largely publicly available information e.g. organizations' websites and publications, traditional and social media articles such as newspapers and blogs, as well as academic literature. Interviews were conducted per phone or internet video call with: OCHA / Libya Crisis Map, Harassmap Egypt, Front for the Defence of Egyptian Protester, Resolve / LRA Crisis Tracker, Safecity India, Agresiones contra periodistas y blogueros en Mexico, and Women's refugee commission / Watchlist on Children.}.
Our research shows that many crowd-sourcing projects
do not inform their sources of the risks entailed with reporting potentially sensitive information using media that can be traced or channels that can be intercepted.
Rarely precautions are taken to secure collected data, in particular sensitive information such as biographic information. In fact, many projects seem to be either unaware of these challenges or when aware they appear ill-equipped to tackle them. 

\afterpage{
\renewcommand*{\arraystretch}{1.2}
\setlength\LTcapwidth{\linewidth}
\begin{footnotesize} 
\begin{longtable}{>{\raggedright\arraybackslash}m{11cm}>{\raggedright\arraybackslash}m{2cm}} \arrayrulecolor{gray!40}
\caption{Crowd-sourcing projects in humanitarian and human rights nature \vspace*{-8pt}} 
\label{tab:projects} \\
\cellcolor{gray!25} \textbf{Platfrom and Description} & \cellcolor{gray!25} \textbf{Technology Used} \\ \hline 

\multicolumn{2}{l}{\textit{AFRICA}} \endhead \hline

\textbf{Libya Crisis Map (Libya)} 

Early March 2011, OCHA activated the Standby Volunteer Task Force (SBTF), a group of volunteers with various field of expertise, to create the Libya Crisis Map to help provide better situational awareness about the situation unfolding on the ground. The UN had no access to the country and OCHA did not have the resources to gather, verify and process the amount of online available information. The website was made available to organizations responding to the Libya crisis with the intention to provide them information that may assist in improving their operational planning.  In April 2011 the SBTF handed over the map to OCHA which maintained it with a group of volunteers until 4 June 2011. 
&
Ushahidi, Skype incl. instant messenger, Google Docs, Google Groups\\ \hline

\textbf{The Front to Defend Egypt Protesters (FDEP) (Egypt)}

Early April 2010, some 34 NGOs in Egypt setup the FDEP to provide legal and informative support to participants to peaceful demonstrations; provide coordinated response to random and mass arrest by the police as well as detention and inhumane treatment of protesters and detainees; and help coordinate efforts and work by human rights groups and lawyers towards their release. By 2011 FDEP's mandate expanded to cover a total of 8 governorates and provide legal support to those tried by military courts. 
To reach this objective, the FDEP set up several hotlines used by activists to report via SMS or calls about arrest, detentions, injuries and the need for lawyers. Protesters facing arrest were given the possibility to SMS their full name, age, ID, health status, time and location of detention. Upon reception of such information FDEP lawyers would go to police stations and other possible places of detentions close to the area of the demonstration to insure that those arrested do not face torture or violence and assist the interrogation process. Lawyers also update a FDEP communication team with the status of the prisoners allowing a medical team to try to gain access in case protesters are injured and a provision committee to provide food, drinks, medication and other necessity. After verification of the information the communication team would post the name of the detainees and their location of detention on the FDEP blog allowing families and relatives to call and provide additional information. The communication team also post media reports in Flickr and Youtube platforms and initiated a Twitter hashtag (\#EgyDefense) to tweet immediate news.  
&
Telephone hotlines, SMS, Twitter, Flickr, Youtube, Blogs, Google maps, Google Drive, RSS feeds \\ \hline

\textbf{Harass map (Egypt and several countries in the world)}

Harassmap was launched in December 2010 by four volunteers using Ushahidi and FrontlineSMS ``with the mission to end the social acceptability of sexual harassment and assault in Egypt''. By means of online and mobile technology, mass media and communications campaigns Harassmap support an on-the-ground mobilization efforts by 700 volunteers spread accross 15 governorates in Egypt focused on changing perceptions so that people start seeing sexual harassment and assault as the crime it actually is and start standing up to it before and when they see it happen. The end goal is to restore a sense of social responsibility and make all of Egypt a ``Safe Zone''.
& Ushahidi, FrontlineSMS, Twitter, Facebook, email
\\ \hline

\textbf{Speak-to-Tweet (Egypt, Syria)}

A joint project by Google and Twitter that in case of Internet blackouts like the one experienced in Egypt begin 2011 and more recently in Syria end 2012 allows users to get their voices heard by calling designated phone numbers and leave a voice message which is automatically tweeted with the hashtag of the country of origin of the call, without the need for an Internet connection.  
&
Twitter, Telephones, Google docs
\\ \hline

\textbf{UN (Somalia and Syria)}

The United Nations reportedly approached the Standby Task Force (SBTF) to pilot the crowd-sourcing of the satellite imagery analysis of a stretch of Somalian territory to tag the possible location of shelters. No operational purposes. Test only.
&
Tommod, Ushahidi 
\\ \hline

\textbf{Voix des Kivus (Democratic Republic of the Congo)}

The objective of this Columbia University project was to examine the potential for using SMS technology to gather conflict event data in real-time using a ``crowd-seeding'' approach instead of a crowd-sourcing approach. Using standard principles of survey research and statistical analysis, 18 sites in the province of South Kivu were sampled. In each site 3 trusted reporters were identified, trained and provided with a mobile phone and reporting instructions. Only they could contribute reports, rather than the crowd with a mobile phone or connection of some sort, as it the case with standard crowd-sourcing platforms.
&
FrontlineSMS, R, LaTeX
\\ \hline
\textbf{LRA Crisis Tracker (CAR, DRC, South Sudan)}

Gather information about the Lord?s Resistance Army movements and attacks from a local early-warning radio network supported by Invisible Children in addition to data sourced from the United Nations, local Non-Government Organizations, and first-hand research to improve efforts to combat LRA atrocities and help communities in need.
To achieve these objectives, Invisible Children deployed HF radios in several location across northern D.R. Congo and larger towns in eastern Central African Republic (CAR) as well as satellite phones to local security committees in more remote and less restive locations in eastern CAR. In Southern Sudan, existing HF radios are informally ``plugged in'' into Invisible Children' radio network. Local security committees gather information about LRA activities and related security incidents from a variety of local sources, both direct and indirect, and transmit them via the HF network or satellite phones to Invisible Children team based in Dungu, D.R. Congo. The Invisible Children staff in Dungu also participate in weekly OCHA led protection cluster as well as the MONUSCO led Joint Information  Operation Cell (JIOC) meetings with other humanitarian actors active in the area who may share information about LRA activities. The collected information is putted into a customized version of the Salesforce customer relationship platform which is ultimately analysed by Resolve analysts who produce regular security briefs. Simultaneously, information about LRA activities for which there is a reasonable doubt that they occurred are published via an interactive map on the LRA Crisis Tracker website.
& 
HF Radios, satellite phones, cloud based Salesforce platform
\\ \hline

\textbf{Hatari (Kenya)}

Allow Nairobi residents to report incidents of crime and corruption in their own voices by SMS, Twitter, smartphone app, email or via the website.
& 
Ushahidi, Frontline SMS, Twitter, Email
\\ \hline

\end{longtable}
\end{footnotesize}
}

\afterpage{
	\renewcommand*{\arraystretch}{1.2}
	\setlength\LTcapwidth{\linewidth}
	\begin{footnotesize} 
		\setcounter{table}{1}
		\begin{longtable}{>{\raggedright\arraybackslash}m{11cm}>{\raggedright\arraybackslash}m{2cm}} \arrayrulecolor{gray!40}
			\caption{Crowd-sourcing projects in humanitarian and human rights nature \vspace*{-8pt}} 
			\label{tab:projects} \\
			\cellcolor{gray!25} \textbf{Platfrom and Description} & \cellcolor{gray!25} \textbf{Technology Used}  \\ \hline 
			
			\multicolumn{2}{l}{\textit{MIDDLE-EAST}} \endhead \hline
			
			\textbf{Syria Tracker (Syria)}
			
			A crowdsourcing effort that has been collecting citizen reports on human rights violations and casualties in Syria, since April 2011 which goals are to provide the number of the fatalities and preserve the name, location and details of each victim. Whenever possible, each name is linked to a photo or video of each casualty. Syria Tracker provides: A continually updated list of eye witness reports from within Syria, often accompanied by media links; aggregate reports including analysis and visualizations of deaths and atrocities in Syria and a stream of content-filtered media from news, social media (Twitter and Facebook) and official sources. 
			&
			Ushahidi, HealthMap platform, Crisis Tracker platform, Twitter, Facebook, Speak-to-Tweet
			\\ \hline
			
			\textbf{Women under siege (Syria)}
			
			Document and map reports of sexual violence in the context of the Syrian conflict discover whether rape and sexual assault are widespread--such evidence can be used to aid the international community in grasping the urgency of what is happening in Syria, and can provide the base for potential future prosecutions. 
			&
			Ushahidi, email, Twitter, app
			\\ \hline
			
			\textbf{Middle East Domestic Help Abuse Reporting (Middle-East)}
			
			Allow human rights organizations, concerned citizens and migrant workers victims of domestic abuse to report alleged incident of abuses to palliate to the lack of a centralised source for data about abuses against migrant workers. 
			&
			Ushahidi crowdmap, email, Twitter, SMS
			\\ \hline
			
		\end{longtable}
	\end{footnotesize}
}

\afterpage{
\renewcommand*{\arraystretch}{1.2}
\setlength\LTcapwidth{\linewidth}
\begin{footnotesize} 
\setcounter{table}{1}
\begin{longtable}{>{\raggedright\arraybackslash}m{11cm}>{\raggedright\arraybackslash}m{2cm}} \arrayrulecolor{gray!40}
\caption{Crowd-sourcing projects in humanitarian and human rights nature \vspace*{-8pt}} 
\label{tab:projects} \\
\cellcolor{gray!25} \textbf{Platfrom and Description} & \cellcolor{gray!25} \textbf{Technology Used}  \\ \hline 

\multicolumn{2}{l}{\textit{AMERICA}} \endhead \hline

\textbf{Hollaback (USA and several countries in the world)}

Expose street harassers by documenting, mapping and sharing incidents of street harassment by means of a smart phone application.
& 
Hollaback app, Google maps
\\ \hline

\textbf{Digital Democracy -- Empowering women in Haiti (Haiti)}

Assist gender-based violence victims and empower women in general in Haiti by means of technology and grass-root activities.
&
Noula (an Ushahidi like platform)
\\ \hline

\textbf{\#PorTodosLosDesaparecidos (Central America)}

To record the 27 thousand missing that the National Human Rights Commission (NHRC) has registered in view of facilitating a direct contact between the victims, citizens, family and the media.
&
Crowdmap, Twitter, Smartphone app, Emails
\\ \hline

\textbf{Agresiones contra periodistas y blogueros en México (Mexico)}

Joint programme of Freedom House, the International Centre for journalists and Mi Mexico to record and map incidents of attacks against journalists, bloggers and citizen reporters in Mexico.
&
Crowdmap, Smartphone app, Twitter, Emails, telephone
\\ \hline

\textbf{Retio (Mexico)}

Crowdsource citizen reports via Twitter about any danger or problems in their areas, including activities of security forces. These reports are recorded in a database and published online to allow users to better understand their environment and let the authorities know that their actions are being monitored with the hope that it will inhibit extortion, arbitrary detention, abuse of authority and police brutality
&
Retio, Twitter
\\ \hline

\multicolumn{2}{l}{\textit{ASIA}} \\ \hline

\textbf{Women Empowerment for Social Change Program (Cambodia)}

Map reported incidents of gender based violence in Cambodia to offers both government officials, key stakeholders and the public the opportunity to track incidences of gender based violence online to increase awareness and work toward immediate intervention and prevention methods. 
&
Ushahidi
\\ \hline

\textbf{Safecity India (India)}

Safecity is an information aggregation platform for victims and witnesses of sexual harassment to report harassment of a sexual nature and help identifying locations where these occurred. The final objective of Safecity is not so much recording information to seek redress for the victims f sexual harassment, but is mainly preventive by highlighting a serious social issue to change the way our society thinks and reacts about sexual harassment, which in time will hopefully lead to a safe and non-violent environment for all. 
&
Ushahidi, Twitter, Email, smart phone app, Interactive Voice Response system
\\ \hline

\multicolumn{2}{l}{\textit{EUROPE}} \\ \hline

\textbf{Rate your Rights Serbia (Serbia)}

An UNDP/UNHCHR initiative in partnership with the Belgrade Centre for Human Rights and media outlet b92 to crowdsource answers to a questionnaire on the state of human rights in Serbia  in the context of the Universal Periodic review and at a later stage to comment on the answers given by the State of Serbia to the same questionnaire.  
&
 Internet based questionnaire 
\\ \hline

\textbf{Istanbul Violence (Turkey)}

Map violence during the June 2013 demonstrations that erupted in Taksim square and spread to other towns across Turkey. 
&
Ushahidi crowd-map, Twitter, emails, app
\\ \hline

\end{longtable}
\end{footnotesize}
}

From a technological point of view, Ushahidi stands out as the most used crowd-sourcing platform in the humanitarian domain. The platform is often deployed without or with minimal customization unless run by professionals. The project teams often lack the necessary resources to use the platform to its full capacity or integrate it with other technologies such as SMS gateways and aggregators or Interactive Voice Response systems. The teams usually settle with the out of the box functionalities without considering which ones need to be deployed to address the needs of the users.
For example, for some NGOs, Internet communication 
security policies prevented access to Ushahidi's website during the Haiti crisis response \citep{Altay2014}. 
In many cases, the dynamic event data aggregated could not be fully integrated into coordination mechanisms because it did not align with the specific information requirements 
of the organisations including large NGOs and the UN \citep{Nelson2010,Morrow2011,Altay2014}. 

A few projects stand out for their professional and innovative approaches. The LRA CrisisTracker project is taking security very seriously and developed a complete codebook to ensure high data quality. It is the only project uncovered insofar that has a methodology in place to assess the veracity of the information gathered. Facing the same challenges as with many other projects, it does fall short in terms of sourcing and does not record enough information to allow for an unbroken chain of custody. Digital Democracy's Empowering Women in Haiti has developed robust feedback loops allowing victims of sexual or gender based violence to seek assistance. The Front to Defend Egyptian Protesters uses a variety of tactics, including crowd-sourcing of information when appropriate. From all the projects reviewed it is the only one that proactively works with lawyers who systematically and personally verify reported information  and follow case files of detained protesters.

In terms of innovation, some developments took place in the sphere of semantic analysis and machine learning with projects such as Syria CrisisTracker, which tried to make sense of publicly available information from the media with the help of algorithms that seek to automatically assess the relevance of published information, clustering these along known topical issues. Other projects which were still on the test bed also showed interesting promises, such as the Standby Task Force aimed to micro-task the analysis of a large piece of satellite imagery in Somalia to help the UN assess the number of displaced persons in a given area. Although these attempts showed shortcomings and needed to be further professionalized, the ideas and concepts that drive them show a promising future if properly used.

\section{Discussion and Future Research}

Information verification is traditionally important and challenging. It becomes even more so with the emergence of new technologies such as social media and crowd-sourcing.  
Due to different traditions, the news and humanitarian organizations have different practices and use different means for information gathering and verification. Both domains could learn from each other's experiences. Although with varied focuses, both tend to expand information verification from a centralized internal approach towards a more community-based external approach, where the content, context and sources are checked by means of e.g. crowd-sourcing and micro-tasking instead of only by the organizations themselves. 
This trend is not without consequences, however. 
Engaged citizens and volunteers who use crowd-sourcing to report incidents and document events, as well as the organizations that deploy crowd-sourcing platforms may come unprepared and inexperienced to the tasks. The potential privacy, safety and security risks and challenges can be overlooked both by the contributors and the organizations. Even when there is awareness, those issues are often challenging to tackle given the time and resources available. 

Beyond warranted privacy, safety and security concerns, the information gathered is often incomplete and is of limited use beyond advocacy. In general, information verification is lacking.
 The basic 5W1H questions (when, where, who did what, why and how) are rarely answered. Many projects developed their own taxonomies that are not well defined, leaving their sources or those manually processing the information guessing what each category refers to. As categories do not align with known International Humanitarian Law and human rights categories, the data gathered could not be easily used or compared unless being recoded.
The quality of the verification procedures, if any, varied largely. Some projects did not attempt to verify the information gathered and published, other applied simplistic methods. For example if the information has been reported by a traditional media it was deemed verified. Some tried to cross-validate the information gathered but did not document or publish their methodologies.
Combined with the frequent lack of information verification, the lack of source evaluation rendered many projects permeable to disinformation attempts which beyond affecting the credibility of the projects could also have real-life effects. 

In terms of scalability and empowerment, many projects which are not run by professionals or supported by established organizations do not adopt a human centred design process when designing and deploying crowd-sourcing technologies; consider desired impact and outcomes, inputs and outputs. Some projects have a pre-determined idea of what could or should be done and hope that by deploying a crowd-sourcing platform people will start reporting information relevant to the project goal. Communication cultures (e.g. prominent use of text or voice) or the level of literacy of the targeted users is rarely taken into consideration. Sensitization efforts around the project are often minimal, relying too often on existing social media channels that are only accessible to a (computer) literate audience. At times, the sole objective is the collection and publication of information, without further use expected. The flow of information remains unidirectional, with no feedback loops that could stimulate the empowerment of the sources and users of the information. The success of such endeavours remain limited, with the crowd-sourcing nature of several projects relying on a few devoted volunteers.

To tackle these challenges, we propose the following research agenda for future work. 
First, the state of the art in the domain of (business, police, civilian and military) intelligence shall be surveyed. There is a rich body of knowledge related to information collection and verification in this domain. 
Limited by time and resources, this project only surveyed and reported on the state of the art in investigative journalism and humanitarian information systems. Interesting research questions include: Can the methodologies and technologies for information collection and verification from investigative journalism and (business, police, civilian and military) intelligence add value to the humanitarian domain? To what extend can those methodologies and technologies be reused and adapted to improve the quality of humanitarian information and meet the needs of crisis decision-making and coordination? 

Second, with respect to information quality for humanitarian responses, clear operational guidelines shall be defined for the desired information quality dimensions, and how to assist achieving the information quality goals during the preparation and information collection stages as well as later stages, regardless whether the information is being crowed-sourced or not. Although there exists a set of humanitarian information management principles\footnote{Theses are: Accessibility, Inclusiveness, Inter-operability, Accountability, Verifiability, Relevance, Impartiality, Humanity, Timeliness, Sustainability, Reliability, Reciprocity and Confidentiality \citep{VandeWalle2008}.}, defining concrete operational guidelines that can be integrated or tailored to the existing practices of major humanitarian actors, is essential to help achieve those goals. Interesting research questions include: How to operationalise the humanitarian information management principles? How to refine or translate those principles into measurable information quality dimensions and metrics so that humantarian actors have well-defined and implementable standards and guidelines to improve information quality during the preparation, information collection and verification stages? 

Third, in the humanitarian domain, resource intensiveness in terms of manpower and time, combined with the complexity of the tasks, are serious impediments to collecting good quality information and to information verification. There have been a number of initiatives and projects that aim to use technology to facilitate those tasks, but the processes and platforms that could effectively and efficiently address the complexity of those tasks are not yet mature. There is an urgent need for technological innovation and breakthrough in the humanitarian domain, but the difficulties and challenges reside not in the technological side or the socio-political side alone but in the combination and dynamics of both. Simply put, the best piece of technology will not be practically useful, especially in crisis situations, if there is a lack of sufficient consideration for privacy, safety and security by design, or if other socio-political context and important values are not taken into careful consideration. With respect to the socio-technical nature of humanitarian information systems, interesting research questions include: 
How to facilitate humanitarian actors to detect and collect relevant information and verify the information using emerging technologies such as smart analytics, machines learning, crowd-sourcing and micro tasking? 
How can such technologies be designed to empower the humanitarian actors and the affected communities alike in a scalable and sustainable manner? 

To summarize, using emerging technologies such as crowd-sourcing and machine learning for information collection and verification is still in its infancy and shows many limitations. There are many challenges ahead calling for research and innovation. Many projects surveyed by this work lacked methodologies and relied heavily on human input for documentation and analysis. Recorded information lacked quality, and taxonomies differed between projects. Methodological standards appeared non-existent. Source evaluation as well as cross-validation was a seldom phenomenon and could  be improved when present. Privacy, safety and security safeguards were rare. Save when projects were accompanied by grass-root activities, feedback loops to affected communities or information contributors had seldom been put in place, restricting empowerment opportunities. Often communication channels required users to have access to the Internet, which limited the user base to a more educated and resourceful crowd, and inhibited the participation and empowerment of less educated and disadvantaged crowds. Unless when the project are run by professionals, project initiators developed few synergies and partnerships with likewise projects and other concerned actors. More coordination, collaboration and knowledge sharing is needed in future projects.

\renewcommand{\bibfont}{\footnotesize}
\printbibliography

@Article{Ackoff1989,
  Title                    = {From data to wisdom},
  Author                   = {R. L. Ackoff},
  Journal                  = {Journal of Applied Systems Analysis},
  Year                     = {1989},
  Pages                    = {3-9},
  Volume                   = {16}
}

@Article{Brandtzaeg2016,
  Title                    = {Emerging Journalistic Verification Practices Concerning Social Media},
  Author                   = {Petter Bae Brandtzaeg and Marika LÃŒders and Jochen Spangenberg and Linda Rath-Wiggins and AsbjÃžrn FÃžlstad},
  Journal                  = {Journalism Practice},
  Year                     = {2016},
  Number                   = {3},
  Pages                    = {323-342},
  Volume                   = {10},

  Doi                      = {10.1080/17512786.2015.1020331},
  Eprint                   = { 
 http://dx.doi.org/10.1080/17512786.2015.1020331
 
},
  Url                      = { 
 http://dx.doi.org/10.1080/17512786.2015.1020331
 
}
}

@PhdThesis{Huang2013,
  Title                    = {Automated Simulation Model Generation},
  Author                   = {Yilin Huang},
  School                   = {Delft Univeristy of Technology},
  Year                     = {2013}
}

@Article{Lillrank2003,
  Title                    = {The quality of information},
  Author                   = {Paul Lillrank},
  Journal                  = {International Journal of Quality and Reliability Management},
  Year                     = {2003},
  Number                   = {6},
  Pages                    = {691 - 703},
  Volume                   = {20}
}

@Book{Loshin2011,
  Title                    = {The Practitioner's Guide to Data Quality Improvement},
  Author                   = {David Loshin},
  Publisher                = {Morgan Kaufmann},
  Year                     = {2011}
}

@Article{Pipino2002,
  Title                    = {Data quality assessment},
  Author                   = {Pipino, Leo L. and Lee, Yang W. and Wang, Richard Y.},
  Journal                  = {Communications of the ACM},
  Year                     = {2002},
  Number                   = {4},
  Pages                    = {211 - 218},
  Volume                   = {45},

  Acmid                    = {506010},
  Address                  = {New York, NY, USA},
  Doi                      = {10.1145/505248.506010},
  ISSN                     = {0001-0782},
  Issue_date               = {April 2002},
  Numpages                 = {8},
  Publisher                = {ACM},
  Url                      = {http://doi.acm.org/10.1145/505248.506010}
}

@Article{Price2005,
  Title                    = {A semiotic information quality framework: development and comparative analysis},
  Author                   = {Rosanne Price and Graeme Shanks},
  Journal                  = {Journal of Information Technology},
  Year                     = {2005},
  Pages                    = {88 - 102},
  Volume                   = {20},
  Doi                      = {10.1057/palgrave.jit.2000038}
}

@Article{Wand1996,
  Title                    = {Anchoring data quality dimensions in ontological foundations},
  Author                   = {Wand, Yair and Wang, Richard Y.},
  Journal                  = {Communications of the ACM},
  Year                     = {1996},
  Number                   = {11},
  Pages                    = {86 - 95},
  Volume                   = {39},

  Acmid                    = {240479},
  Address                  = {New York, NY, USA},
  Doi                      = {10.1145/240455.240479},
  ISSN                     = {0001-0782},
  Issue_date               = {Nov. 1996},
  Numpages                 = {10},
  Publisher                = {ACM},
  Url                      = {http://doi.acm.org/10.1145/240455.240479}
}

@InProceedings{Ozdamar2015,
  Title                    = {{Models, solutions and enabling technologies in humanitarian logistics}},
  Author                   = {{\"{O}}zdamar, Linet and Ertem, Mustafa Alp},
  Booktitle                = {European Journal of Operational Research},
  Year                     = {2015},
  Month                    = {jul},
  Number                   = {1},
  Pages                    = {55--65},
  Publisher                = {North-Holland},
  Volume                   = {244},
  Doi                      = {10.1016/j.ejor.2014.11.030},
  ISBN                     = {0377-2217},
  ISSN                     = {03772217},
  Url                      = {https://www.sciencedirect.com/science/article/pii/S0377221714009539}
}

@Article{Ali2017,
  Title                    = {{Rule-guided human classification of Volunteered Geographic Information}},
  Author                   = {Ali, Ahmed Loai and Falomir, Zoe and Schmid, Falko and Freksa, Christian},
  Journal                  = {ISPRS Journal of Photogrammetry and Remote Sensing},
  Year                     = {2017},

  Month                    = {may},
  Pages                    = {3--15},
  Volume                   = {127},
  Doi                      = {10.1016/j.isprsjprs.2016.06.003},
  ISSN                     = {09242716},
  Publisher                = {Elsevier},
  Url                      = {https://www.sciencedirect.com/science/article/pii/S0924271616301137}
}

@Book{alpaydin2014introduction,
  Title                    = {Introduction to machine learning},
  Author                   = {Alpaydin, Ethem},
  Publisher                = {MIT press},
  Year                     = {2014}
}

@Article{Altay2014,
  Title                    = {Challenges in humanitarian information management and exchange: evidence from Haiti},
  Author                   = {Nezih Altay and Melissa Labonte},
  Journal                  = {Disasters},
  Year                     = {2014},
  Number                   = {s1},
  Pages                    = {50-72},
  Volume                   = {38},
  Doi                      = {10.1111/disa.12052},
  Eprint                   = {https://onlinelibrary.wiley.com/doi/pdf/10.1111/disa.12052},
  Url                      = {https://onlinelibrary.wiley.com/doi/abs/10.1111/disa.12052}
}

@Article{Anson2017,
  Title                    = {{Analysing social media data for disaster preparedness: Understanding the opportunities and barriers faced by humanitarian actors}},
  Author                   = {Anson, Susan and Watson, Hayley and Wadhwa, Kush and Metz, Karin},
  Journal                  = {International Journal of Disaster Risk Reduction},
  Year                     = {2017},

  Month                    = {mar},
  Pages                    = {131--139},
  Volume                   = {21},
  Doi                      = {10.1016/j.ijdrr.2016.11.014},
  ISBN                     = {2212-4209},
  ISSN                     = {22124209},
  Publisher                = {Elsevier},
  Url                      = {https://www.sciencedirect.com/science/article/pii/S221242091630156X}
}

@InProceedings{Basu2016,
  Title                    = {{Post disaster situation awareness and decision support through interactive crowdsourcing}},
  Author                   = {Basu, Moumita and Bandyopadhyay, Somprakash and Ghosh, Saptarshi},
  Booktitle                = {Procedia Engineering},
  Year                     = {2016},
  Month                    = {jan},
  Pages                    = {167--173},
  Publisher                = {Elsevier},
  Volume                   = {159},
  Doi                      = {10.1016/j.proeng.2016.08.151},
  ISBN                     = {18777058 (ISSN)},
  ISSN                     = {18777058},
  Url                      = {https://www.sciencedirect.com/science/article/pii/S1877705816322998}
}

@TechReport{Bruno2011,
  Title                    = {Tweet first, verify later: How real-time information is changing the
coverage of worldwide crisis events},
  Author                   = {Nicola Bruno},
  Institution              = {Reuters Institute for the Study of Journalism, University of Oxford},
  Year                     = {2011},
  Note                     = {Thomson Reuters Foundation},

  HowPublished             = {Reuters Institute for the Study of Journalism}
}

@Article{Callaghan2016,
  Title                    = {{Disaster management, crowdsourced R{\&}D and probabilistic innovation theory: Toward real time disaster response capability}},
  Author                   = {Callaghan, Christian William},
  Journal                  = {International Journal of Disaster Risk Reduction},
  Year                     = {2016},

  Month                    = {aug},
  Pages                    = {238--250},
  Volume                   = {17},
  Doi                      = {10.1016/j.ijdrr.2016.05.004},
  ISSN                     = {22124209},
  Publisher                = {Elsevier},
  Url                      = {https://www.sciencedirect.com/science/article/pii/S2212420915300698}
}

@Article{Carley2016,
  Title                    = {{Crowd sourcing disaster management: The complex nature of Twitter usage in Padang Indonesia}},
  Author                   = {Carley, Kathleen M. and Malik, Momin and Landwehr, Peter M. and Pfeffer, J{\"{u}}rgen and Kowalchuck, Michael},
  Journal                  = {Safety Science},
  Year                     = {2016},

  Month                    = {mar},
  Pages                    = {48--61},
  Volume                   = {90},
  Doi                      = {10.1016/j.ssci.2016.04.002},
  ISBN                     = {09257535},
  ISSN                     = {18791042}
}

@Article{Castillo2013,
  Title                    = {{Predicting information credibility in time-sensitive social media}},
  Author                   = {Castillo, Carlos and Mendoza, Marcelo and Poblete, Barbara},
  Journal                  = {Internet Research},
  Year                     = {2013},

  Month                    = {oct},
  Number                   = {5},
  Pages                    = {560--588},
  Volume                   = {23},
  Doi                      = {10.1108/IntR-05-2012-0095},
  Editor                   = {{Gayo-Avello, Panagiotis Takis Metax}, Daniel},
  ISBN                     = {1066-2243},
  ISSN                     = {1066-2243},
  Url                      = {http://www.emeraldinsight.com/doi/10.1108/IntR-05-2012-0095}
}

@InProceedings{Cobb2014,
  Title                    = {Designing for the Deluge: Understanding \& Supporting the Distributed, Collaborative Work of Crisis Volunteers},
  Author                   = {Cobb, Camille and McCarthy, Ted and Perkins, Annuska and Bharadwaj, Ankitha and Comis, Jared and Do, Brian and Starbird, Kate},
  Booktitle                = {Proceedings of the 17th ACM Conference on Computer Supported Cooperative Work \& Social Computing},
  Year                     = {2014},

  Address                  = {New York, NY, USA},
  Pages                    = {888-899},
  Publisher                = {ACM},
  Series                   = {CSCW '14},

  Acmid                    = {2531712},
  Doi                      = {10.1145/2531602.2531712},
  ISBN                     = {978-1-4503-2540-0},
  Location                 = {Baltimore, Maryland, USA},
  Numpages                 = {12},
  Url                      = {http://doi.acm.org/10.1145/2531602.2531712}
}

@Article{Conrado2016,
  Title                    = {{Managing social media uncertainty to support the decision making process during Emergencies}},
  Author                   = {Conrado, Silvia Planella and Neville, Karen and Woodworth, Simon and O'Riordan, Sheila},
  Journal                  = {Journal of Decision Systems},
  Year                     = {2016},
  Number                   = {sup1},
  Pages                    = {171-181},
  Volume                   = {25},
  Doi                      = {10.1080/12460125.2016.1187396},
  ISBN                     = {1166-8636},
  ISSN                     = {21167052},
  Publisher                = {Taylor {\&} Francis},
  Url                      = {https://doi.org/10.1080/12460125.2016.1187396}
}

@TechReport{Coyle2009,
  Title                    = {New Technologies in Emergencies and Conflicts: The Role of Information and Social Networks},
  Author                   = {Coyle, D. and Meier, P.},
  Institution              = {United Nations Foundation \& Vodafone Foundation},
  Year                     = {2009}
}

@Article{crowley2013decision,
  Title                    = {Decision support using linked, social, and sensor data},
  Author                   = {Crowley, David N and Dabrowski, Maciej and Breslin, John G},
  Year                     = {2013}
}

@Article{DAUME20149,
  Title                    = {Forest monitoring and social media – Complementary data sources for ecosystem surveillance?},
  Author                   = {Stefan Daume and Matthias Albert and Klaus von Gadow},
  Journal                  = {Forest Ecology and Management},
  Year                     = {2014},
  Note                     = {Forest Observational Studies: “Data Sources for Analysing Forest Structure and Dynamics”},
  Pages                    = {9 - 20},
  Volume                   = {316},

  Doi                      = {https://doi.org/10.1016/j.foreco.2013.09.004},
  ISSN                     = {0378-1127},
  Url                      = {http://www.sciencedirect.com/science/article/pii/S037811271300618X}
}

@InProceedings{Diakopoulos2012,
  Title                    = {Finding and assessing social media information sources in the context of journalism},
  Author                   = {Diakopoulos, Nicholas and {De Choudhury}, Munmun and Naaman, Mor},
  Booktitle                = {Proceedings of the 2012 ACM annual conference on Human Factors in Computing Systems - CHI '12},
  Year                     = {2012},

  Address                  = {New York, NY, USA},
  Pages                    = {2451},
  Publisher                = {ACM},
  Series                   = {CHI '12},
  Doi                      = {10.1145/2207676.2208409},
  ISBN                     = {9781450310154},
  Url                      = {http://doi.acm.org/10.1145/2207676.2208409 http://dl.acm.org/citation.cfm?doid=2207676.2208409}
}

@InProceedings{Dugdale2012,
  Title                    = {Social Media and SMS in the Haiti Earthquake},
  Author                   = {Dugdale, Julie and Van de Walle, Bartel and Koeppinghoff, Corinna},
  Booktitle                = {Proceedings of the 21st International Conference on World Wide Web},
  Year                     = {2012},

  Address                  = {New York, NY, USA},
  Pages                    = {713--714},
  Publisher                = {ACM},
  Series                   = {WWW '12 Companion},

  Acmid                    = {2188189},
  Doi                      = {10.1145/2187980.2188189},
  ISBN                     = {978-1-4503-1230-1},
  Location                 = {Lyon, France},
  Numpages                 = {2},
  Url                      = {http://doi.acm.org/10.1145/2187980.2188189}
}

@Article{Gao2011,
  Title                    = {{Harnessing the crowdsourcing power of social media for disaster relief}},
  Author                   = {Gao, Huiji and Barbier, Geoffrey and Goolsby, Rebecca},
  Journal                  = {IEEE Intelligent Systems},
  Year                     = {2011},
  Number                   = {3},
  Pages                    = {10-14},
  Volume                   = {26},
  Address                  = {Piscataway, NJ, USA},
  Doi                      = {10.1109/MIS.2011.52},
  ISBN                     = {1541-1672},
  ISSN                     = {15411672},
  Publisher                = {IEEE Educational Activities Department},
  Url                      = {http://dx.doi.org/10.1109/MIS.2011.52}
}

@Article{Harrison2010,
  Title                    = {{User-Generated Content and Gatekeeping At the BBc Hub}},
  Author                   = {Harrison, J.},
  Journal                  = {Journalism
Studies},
  Year                     = {2010},
  Number                   = {2},
  Pages                    = {243–256},
  Volume                   = {11}
}

@Article{Haworth2016,
  Title                    = {{Emergency management perspectives on volunteered geographic information: Opportunities, challenges and change}},
  Author                   = {Haworth, Billy},
  Journal                  = {Computers, Environment and Urban Systems},
  Year                     = {2016},

  Month                    = {may},
  Pages                    = {189--198},
  Volume                   = {57},
  Doi                      = {10.1016/J.COMPENVURBSYS.2016.02.009},
  ISSN                     = {0198-9715},
  Publisher                = {Pergamon},
  Url                      = {https://www.sciencedirect.com/science/article/pii/S0198971516300175}
}

@TechReport{Heinzelman2010,
  Title                    = {Crowdsourcing Crisis 
Information in Disaster - Affected Haiti},
  Author                   = {Jessica Heinzelman and Carol Waters},
  Institution              = {United States Institute Of Peace (www.usip.org)},
  Year                     = {2010},

  Address                  = {1200 17th Street NW, Washington, DC 2003}
}

@InCollection{Hermida2015,
  Title                    = {Nothing But the Truth: Redrafting the Journalistic Boundary of
Verification},
  Author                   = {Alfred Hermida},
  Booktitle                = {Boundaries of Journalism: Professionalism, Practices and Participation},
  Publisher                = {Taylor \& Frances},
  Year                     = {2015},
  Chapter                  = {2},
  Editor                   = {Matt Carlson and Seth Lewis},
  Number                   = {37-50},

  Journal                  = {Boundaries of Journalism}
}

@Article{Hermida2014,
  Title                    = {Sourcing the Arab spring: A case study of
Andy Carvin's sources on twitter during the Tunisian and Egyptian revolutions},
  Author                   = {Hermida, A. and Lewis, S. C. and Zamith, R.},
  Journal                  = {Journal of Computer-Mediated Communication},
  Year                     = {2014},
  Number                   = {3},
  Pages                    = {479-499},
  Volume                   = {09}
}

@InCollection{Howe2008,
  Title                    = {The Wisdom of the Crowd Resides in How the Crowd is Used},
  Author                   = {Jeff Howe},
  Booktitle                = {Nieman Reports - The Search for True North: New Directions in a New Territory},
  Publisher                = {Niemand Foundation at Harvard University},
  Year                     = {2008},
  Number                   = {62},
  Pages                    = {47-50},
  Volume                   = {4},

  HowPublished             = {http://niemanreports.org/articles/the-wisdom-of-the-crowd-resides-in-how-the-crowd-is-used/},
  Institution              = {Nieman Reports}
}

@Article{Hung2016,
  Title                    = {{Methods for assessing the credibility of volunteered geographic information in flood response: A case study in Brisbane, Australia}},
  Author                   = {Hung, Kuo-Chih and Kalantari, Mohsen and Rajabifard, Abbas},
  Journal                  = {Applied Geography},
  Year                     = {2016},

  Month                    = {mar},
  Pages                    = {37--47},
  Volume                   = {68},
  Doi                      = {10.1016/j.apgeog.2016.01.005},
  ISBN                     = {0143-6228},
  ISSN                     = {01436228},
  Publisher                = {Pergamon},
  Url                      = {https://www.sciencedirect.com/science/article/pii/S0143622816300054 http://linkinghub.elsevier.com/retrieve/pii/S0143622816300054}
}

@InProceedings{Kang2012,
  Title                    = {{Modeling topic specific credibility on twitter}},
  Author                   = {Kang, Byungkyu and O'Donovan, John and H{\"{o}}llerer, Tobias},
  Booktitle                = {Proceedings of the 2012 ACM international conference on Intelligent User Interfaces - IUI '12},
  Year                     = {2012},

  Address                  = {New York, NY, USA},
  Pages                    = {179},
  Publisher                = {ACM},
  Series                   = {IUI '12},
  Doi                      = {10.1145/2166966.2166998},
  ISBN                     = {9781450310482},
  Url                      = {http://doi.acm.org/10.1145/2166966.2166998 http://dl.acm.org/citation.cfm?doid=2166966.2166998}
}

@Article{lin2016,
  Title                    = {Toward Better Data Veracity in Mobile Cloud Computing: A Context-Aware and Incentive-Based Reputation Mechanism},
  Author                   = {Lin, Hui and Hu, Jia and Tian, Youliang and Yang, Li and Xu, Li},
  Year                     = {2016},

  Month                    = {12},
  Volume                   = {387},

  Booktitle                = {Information Sciences}
}

@Article{Liu2016,
  Title                    = {Reuters Tracer: A Large Scale System of Detecting {\&} Verifying Real-Time News Events from Twitter},
  Author                   = {Liu, Xiaomo and Li, Quanzhi and Nourbakhsh, Armineh and Fang, Rui and Thomas, Merine and Anderson, Kajsa and Kociuba, Russ and Vedder, Mark and Pomerville, Steven and Wudali, Ramdev and Martin, Robert and Duprey, John and Vachher, Arun and Keenan, William and Research, Sameena Shah and Reuters, Development Thomson},
  Journal                  = {CIKM},
  Year                     = {2016},
  Pages                    = {207--216},
  Doi                      = {10.1145/2983323.2983363},
  ISBN                     = {978-1-4503-4073-1}
}

@Article{Ludwig2017,
  Title                    = {{Situated crowdsourcing during disasters: Managing the tasks of spontaneous volunteers through public displays}},
  Author                   = {Ludwig, Thomas and Kotthaus, Christoph and Reuter, Christian and van Dongen, S{\"{o}}ren and Pipek, Volkmar},
  Journal                  = {International Journal of Human Computer Studies},
  Year                     = {2017},

  Month                    = {jun},
  Pages                    = {103--121},
  Volume                   = {102},
  Doi                      = {10.1016/j.ijhcs.2016.09.008},
  ISBN                     = {10715819 (ISSN)},
  ISSN                     = {10959300},
  Publisher                = {Academic Press},
  Url                      = {https://www.sciencedirect.com/science/article/pii/S1071581916301197}
}

@Article{Martin2016,
  Title                    = {Information Verification in the Digital Age: The News Library Perspective},
  Author                   = {Martin, Nora},
  Year                     = {2016},

  Month                    = {07},
  Pages                    = {i-51},
  Volume                   = {2},

  Booktitle                = {Synthesis Lectures on Emerging Trends in Librarianship}
}

@Book{Meier2015,
  Title                    = {Digital Humanitarians: How Big Data Is Changing the Face of Humanitarian Response},
  Author                   = {Patrick Meier},
  Publisher                = {CRC Press},
  Year                     = {2015}
}

@Article{Meier2011,
  Title                    = {{New information technologies and their impact on the humanitarian sector}},
  Author                   = {Meier, Patrick},
  Journal                  = {International Review of the Red Cross},
  Year                     = {2011},
  Number                   = {884},
  Pages                    = {1239-1263},
  Volume                   = {93},
  Booktitle                = {International Review of the Red Cross},
  Doi                      = {10.1017/S1816383112000318},
  ISBN                     = {1816383112},
  ISSN                     = {18163831},
  Publisher                = {Cambridge University Press},
  Url                      = {http://www.journals.cambridge.org/abstract{\_}S1816383112000318}
}

@Book{michalski2013machine,
  Title                    = {Machine learning: An artificial intelligence approach},
  Author                   = {Michalski, Ryszard S and Carbonell, Jaime G and Mitchell, Tom M},
  Publisher                = {Springer Science, Business Media},
  Year                     = {2013}
}

@TechReport{Morrow2011,
  Title                    = {Independent Evaluation of the Ushahidi Haiti Project. Development Information 
Systems International Ushahidi Haiti Project},
  Author                   = {Nathan Morrow and Nancy Mock  and Adam Papendieck  and Nicholas Kocmich},
  Institution              = {Ushahidi},
  Year                     = {2011}
}

@TechReport{Nelson2010,
  Title                    = {Media, Information Systems and Communities: Lessons from Haiti},
  Author                   = {Anne Nelson and Ivan Sigal and Dean Zambrano},
  Institution              = {John S. and James L. Knight Foundation},
  Year                     = {2010}
}

@TechReport{Newman2009,
  Title                    = {The rise of social media and its impact on mainstream journalism: A study of how newspapers and broadcasters in the UK and
US are responding to a wave of participatory social media,
and a historic shift in control towards individual consumers},
  Author                   = {Nic Newman},
  Institution              = {Reuters Institute for the Study of Journalism, University of Oxford},
  Year                     = {2009},

  HowPublished             = {Reuters Institute for the Study of Journalism}
}

@Article{Norheim2010,
  Title                    = {{Crowdsourcing for Crisis Mapping in Haiti}},
  Author                   = {Norheim-Hagtun, Ida and Meier, Patrick},
  Journal                  = {Innovations: Technology, Governance, Globalization},
  Year                     = {2010},

  Month                    = {mar},
  Pages                    = {81-89},
  Volume                   = {5},
  Doi                      = {10.1162/INOV_a_00046},
  ISBN                     = {10.1162/INOV{\_}a{\_}00046},
  ISSN                     = {1558-2477},
  Url                      = {http://www.mitpressjournals.org/doi/pdf/10.1162/INOV{\_}a{\_}00046}
}

@Article{Norris2017,
  author    = {Wendy Norris},
  journal   = {Journalism Practice},
  title     = {Digital Humanitarians: Citizen journalists on the virtual front line of natural and human-caused disasters},
  year      = {2017},
  number    = {2-3},
  pages     = {213-228},
  volume    = {11},
  doi       = {10.1080/17512786.2016.1228471},
  eprint    = {https://doi.org/10.1080/17512786.2016.1228471},
  publisher = {Routledge},
  url       = {https://doi.org/10.1080/17512786.2016.1228471},
}

@Booklet{OCHA2006,
  Title                    = {Guidelines for OCHA Field Information Management},

  Address                  = {New York: United Nations},
  Author                   = {OCHA},
  Year                     = {2006}
}

@Article{Okolloh2009,
  Title                    = {Ushahidi, or `testimony': Web 2.0 
tools for crowdsourcing crisis
information},
  Author                   = {Ory Okolloh},
  Journal                  = {Participatory Learning and Action},
  Year                     = {2009},
  Pages                    = {65-70},
  Volume                   = {59}
}

@Article{Panagiotopoulos2016,
  Title                    = {{Social media in emergency management: Twitter as a tool for communicating risks to the public}},
  Author                   = {Panagiotopoulos, Panos and Barnett, Julie and Bigdeli, Alinaghi Ziaee and Sams, Steven},
  Journal                  = {Technological Forecasting and Social Change},
  Year                     = {2016},

  Month                    = {oct},
  Pages                    = {86--96},
  Volume                   = {111},
  Doi                      = {10.1016/j.techfore.2016.06.010},
  ISBN                     = {0040162516},
  ISSN                     = {00401625},
  Publisher                = {North-Holland},
  Url                      = {https://www.sciencedirect.com/science/article/pii/S0040162516301196}
}

@Article{Pantti2015,
  Title                    = {The Fragility  of the Photo-Truth: Verification  of Amateur  Images in Finnish Newsrooms},
  Author                   = {Mervi Pantti and Stefanie Siren},
  Journal                  = {Digital 
Journalism},
  Year                     = {2015},
  Number                   = {4},
  Volume                   = {3}
}

@InProceedings{Popoola2013,
  Title                    = {Information Verification During Natural Disasters},
  Author                   = {Popoola, Abdulfatai and Krasnoshtan, Dmytro and Toth, Attila-Peter and Naroditskiy, Victor and Castillo, Carlos and Meier, Patrick and Rahwan, Iyad},
  Booktitle                = {Proceedings of the 22Nd International Conference on World Wide Web},
  Year                     = {2013},

  Address                  = {New York, NY, USA},
  Pages                    = {1029--1032},
  Publisher                = {ACM},
  Series                   = {WWW '13 Companion},

  Acmid                    = {2488111},
  Doi                      = {10.1145/2487788.2488111},
  ISBN                     = {978-1-4503-2038-2},
  Location                 = {Rio de Janeiro, Brazil},
  Numpages                 = {4},
  Url                      = {http://doi.acm.org/10.1145/2487788.2488111}
}

@Article{Riccardi2016,
  Title                    = {{The power of crowdsourcing in disaster response operations}},
  Author                   = {Riccardi, Mark T.},
  Journal                  = {International Journal of Disaster Risk Reduction},
  Year                     = {2016},

  Month                    = {dec},
  Pages                    = {123--128},
  Volume                   = {20},
  Doi                      = {10.1016/j.ijdrr.2016.11.001},
  ISSN                     = {22124209},
  Publisher                = {Elsevier},
  Url                      = {https://www.sciencedirect.com/science/article/pii/S2212420916302199}
}

@Article{Rogstadius2013,
  Title                    = {CrisisTracker: Crowdsourced social media curation for disaster awareness},
  Author                   = {J. Rogstadius and M. Vukovic and C. A. Teixeira and V. Kostakos and E. Karapanos and J. A. Laredo},
  Journal                  = {IBM Journal of Research and Development},
  Year                     = {2013},

  Month                    = {Sept},
  Number                   = {5},
  Pages                    = {4:1-4:13},
  Volume                   = {57},

  Doi                      = {10.1147/JRD.2013.2260692},
  ISSN                     = {0018-8646}
}

@Article{Schifferes2014,
  Title                    = {{Identifying and Verifying News through Social Media: Developing a user-centred tool for professional journalists}},
  Author                   = {Schifferes, Steve and Newman, Nic and Thurman, Neil and Corney, David and G{\"{o}}ker, Ayse and Martin, Carlos},
  Journal                  = {Digital Journalism},
  Year                     = {2014},

  Month                    = {jul},
  Number                   = {3},
  Pages                    = {406--418},
  Volume                   = {2},
  Doi                      = {10.1080/21670811.2014.892747},
  ISBN                     = {Schifferes, S., Newman, N., Thurman, N. {\textless}http://openaccess.city.ac.uk/view/creators{\_}id/n=2Ej=2Ethurman.html{\textgreater}, Corney, D., Goker, A.S. {\&} Martin, C. (2014). Identifying and verifying news through social media: Developing a user-centred tool for professional journalists. Digital Journalism, doi: 10.1080/21670811.2014.892747 {\textless}http://dx.doi.org/10.1080/21670811.2014.892747{\textgreater}},
  ISSN                     = {2167082X},
  Url                      = {http://www.tandfonline.com/doi/abs/10.1080/21670811.2014.892747}
}

@Article{shamala2017integrating,
  Title                    = {Integrating information quality dimensions into information security risk management (ISRM)},
  Author                   = {Shamala, Palaniappan and Ahmad, Rabiah and Zolait, Ali and Sedek, Muliati},
  Journal                  = {Journal of Information Security and Applications},
  Year                     = {2017},
  Pages                    = {1--10},
  Volume                   = {36},

  Publisher                = {Elsevier}
}

@Article{SHANTEAU199275,
  Title                    = {How much information does an expert use? Is it relevant?},
  Author                   = {James Shanteau},
  Journal                  = {Acta Psychologica},
  Year                     = {1992},
  Number                   = {1},
  Pages                    = {75 - 86},
  Volume                   = {81},

  Doi                      = {https://doi.org/10.1016/0001-6918(92)90012-3},
  ISSN                     = {0001-6918},
  Url                      = {http://www.sciencedirect.com/science/article/pii/0001691892900123}
}

@InCollection{Soden2014,
  Title                    = {{From Crowdsourced Mapping to Community Mapping: The Post earthquake Work of OpenStreetMap Haiti}},
  Author                   = {Soden, Robert and Palen, Leysia},
  Booktitle                = {COOP 2014 - Proceedings of the 11th International Conference on the Design of Cooperative Systems},
  Publisher                = {Springer International Publishing},
  Year                     = {2014},

  Address                  = {Cham},
  Editor                   = {Rossitto, Chiara and Ciolfi, Luigina and Martin, David and Conein, Bernard},
  Pages                    = {311-326},
  Doi                      = {10.1007/978-3-319-06498-7_19},
  ISBN                     = {978-3-319-06497-0},
  Url                      = {http://link.springer.com/10.1007/978-3-319-06498-7{\_}19}
}

@Article{Spence2016,
  Title                    = {{Social media and crisis research: Data collection and directions}},
  Author                   = {Spence, Patric R. and Lachlan, Kenneth A. and Rainear, Adam M.},
  Journal                  = {Computers in Human Behavior},
  Year                     = {2016},

  Month                    = {jan},
  Pages                    = {667--672},
  Volume                   = {54},
  Doi                      = {10.1016/j.chb.2015.08.045},
  ISBN                     = {0747-5632},
  ISSN                     = {07475632},
  Publisher                = {Pergamon},
  Url                      = {https://www.sciencedirect.com/science/article/pii/S0747563215005270}
}

@Article{Sump2012,
  Title                    = {{Making the Most of Twitter}},
  Author                   = {Sump-Crethar, A Nicole},
  Journal                  = {Reference Librarian},
  Year                     = {2012},
  Number                   = {4},
  Pages                    = {349--354},
  Volume                   = {53},
  Doi                      = {10.1080/02763877.2012.704566},
  ISSN                     = {02763877},
  Publisher                = {Routledge},
  Url                      = {https://doi.org/10.1080/02763877.2012.704566}
}

@Article{Takahashi2015,
  Title                    = {{Communicating on Twitter during a disaster: An analysis of tweets during Typhoon Haiyan in the Philippines}},
  Author                   = {Takahashi, Bruno and Tandoc, Edson C. and Carmichael, Christine},
  Journal                  = {Computers in Human Behavior},
  Year                     = {2015},

  Month                    = {sep},
  Pages                    = {392--398},
  Volume                   = {50},
  Doi                      = {10.1016/j.chb.2015.04.020},
  ISBN                     = {0747-5632},
  ISSN                     = {07475632},
  Publisher                = {Pergamon},
  Url                      = {https://www.sciencedirect.com/science/article/pii/S0747563215003076}
}

@InProceedings{Tapia2011,
  Title                    = {Seeking    the    Trustworthy    Tweet:    Can    Microblogged    
Data    Fit    the    Information    Needs    of    Disaster    Response    and    Humanitarian    Relief    Organizations},
  Author                   = {Tapia, A. and
    Bajpai,    K. and   Jansen,    B.J. and  Yen,    J.},
  Booktitle                = {Proceedings of the 8th International    Conference    on Information    Systems    for    Crisis    Response    and    Management
    (ISCRAM)},
  Year                     = {2011},

  Address                  = {Lisbon, Portugal}
}

@InProceedings{VandeWalle2008,
  Title                    = {Humanitarian Information
Management and Systems},
  Author                   = {Van de Walle, B. and Van Den Eede, G. and Muhren, W.},
  Booktitle                = {Second International Workshop on Mobile Information Technology for Emergency Response, Mobile Response 2008},
  Year                     = {2008},
  Number                   = {12-21},
  Publisher                = {Springer Berlin Heidelberg}
}

@InProceedings{Walton2011,
  Title                    = {Defining ``fast'': Factors affecting the experience of speed in humanitarian logistics},
  Author                   = {Walton, R. and Mays, R. and Haselkorn, M.},
  Booktitle                = {Proceedings of the 8th International    Conference    on Information    Systems    for    Crisis    Response    and    Management
    (ISCRAM)},
  Year                     = {2011},

  Address                  = {Lisbon, Portugal}
}

@InCollection{Wardle2014,
  Title                    = {Verifying User-Generated Content},
  Author                   = {Claire Wardle},
  Booktitle                = {Verification Handbook: An ultimate guide on digital age sourcing for
emergency coverage},
  Publisher                = {The European Journalism Centre},
  Year                     = {2014},
  Editor                   = {Silverman, C.},

  HowPublished             = {European Journalism Centre}
}

@Book{yildiz2015handbook,
  Title                    = {Handbook of research on media literacy in the digital age},
  Author                   = {Yildiz, Melda N},
  Publisher                = {IGI Global},
  Year                     = {2015}
}

@Misc{Yuan2018,
  Title                    = {{Feasibility study of using crowdsourcing to identify critical affected areas for rapid damage assessment: Hurricane Matthew case study}},

  Author                   = {Yuan, Faxi and Liu, Rui},
  Month                    = {feb},
  Year                     = {2018},
  Booktitle                = {International Journal of Disaster Risk Reduction},
  Doi                      = {10.1016/j.ijdrr.2018.02.003},
  ISSN                     = {22124209},
  Publisher                = {Elsevier},
  Url                      = {https://www.sciencedirect.com/science/article/pii/S221242091830150X}
}

@Article{Ziemke2012,
  author    = {Jen Ziemke},
  journal   = {Journal of Map \& Geography Libraries},
  title     = {Crisis Mapping: The Construction of a New Interdisciplinary Field?},
  year      = {2012},
  number    = {2},
  pages     = {101-117},
  volume    = {8},
  doi       = {10.1080/15420353.2012.662471},
  eprint    = {https://doi.org/10.1080/15420353.2012.662471},
  publisher = {Routledge},
  url       = {https://doi.org/10.1080/15420353.2012.662471},
}

@Book{Silverman2014,
  Title                    = {Verification Handbook: An ultimate guide on digital age sourcing for
emergency coverage},
  Editor                   = {Silverman, C.},
  Publisher                = {The European Journalism Centre},
  Year                     = {2014},

  HowPublished             = {European Journalism Centre}
}

@MastersThesis{Vaporidis2019,
  author = {Dimitrios-Marios Vaporidis},
  school = {Delft University of Technology},
  title  = {Detecting Rumors in Twitter for Humanitarian Activities},
  year   = {2019},
}

\end{document}